%% file: main.tex
\documentclass[leqno]{article}

\usepackage[letterpaper]{geometry}
\usepackage{tcolorbox}

\usepackage{amsmath}
\usepackage{amssymb}
\usepackage{mathtools}
\usepackage{ltexpprt}
\usepackage{hyperref}
\hypersetup{breaklinks=true, colorlinks=true, linkcolor={blue!90!black}, citecolor={blue!90!black}, urlcolor={blue!90!black}}
\usepackage{doi}
\usepackage[nameinlink]{cleveref}
\crefname{code}{Listing}{Listings}
\Crefname{code}{Listing}{Listings}

\usepackage{microtype}
\usepackage{graphicx}
\usepackage{booktabs} %

\usepackage[newfloat,frozencache,cachedir=minted-cache]{minted}
\usepackage{braket}  %
\usepackage{multicol}
\DeclareFloatingEnvironment[fileext=cd,placement={!ht},name=Listing]{code}
\usepackage[textsize=tiny]{todonotes}

\newcommand{\matt}[1]{}

\begin{document}

\newcommand\relatedversion{}
\renewcommand\relatedversion{} %

\title{\Large Challenges with Differentiable Quantum Dynamics\relatedversion}
\author{Sri Hari Krishna Narayanan \thanks{Argonne National Laboratory, Mathematics and Computer Division, Lemont, IL 60439} \thanks{Corresponding Author: \texttt{snarayan@mcs.anl.gov}}
\and Michael A. Perlin \thanks{Infleqtion, Inc., 141 West Jackson Blvd Suite 1875 Chicago, IL 60604}
\and Robert Lewis-Swan \thanks{Center for Quantum Research and Technology, Homer L.~Dodge Department of Physics and Astronomy, The University of Oklahoma, Norman, Oklahoma 73019}
\and Jeffrey Larson \footnotemark[1]
\and Matt Menickelly \footnotemark[1]
\and Jan H\"uckelheim \footnotemark[1]
\and Paul Hovland \footnotemark[1]}

\date{}

\maketitle

\begin{abstract} \small\baselineskip=9pt 
Differentiable quantum dynamics require automatic differentiation of a complex-valued initial value problem, which numerically integrates a system of ordinary differential equations from a specified initial condition, as well as the eigendecomposition of a matrix.
We explored several automatic differentiation frameworks for these tasks, finding that no framework natively supports our application requirements. 
We therefore demonstrate a need for broader support of complex-valued, differentiable numerical integration in scientific computing libraries.
\end{abstract}

\section{Introduction}
\label{sec:introduction}
Quantum computing leverages quantum-mechanical phenomena  such as superposition and entanglement to perform computations that are inaccessible to a classical computer.
Quantum sensing similarly leverages quantum phenomena to surpass classical limits on measurement precision with a fixed amount of resources.
In a closed (isolated) quantum system, the state of $n$ quantum bits (qubits) can be described by a  \emph{pure state}, which is represented by a unit vector in a $2^n$-dimensional complex vector space.
Quantum logic gates for quantum computation are typically represented by a $2^n \times 2^n$ unitary matrix acting on the underlying vector space of pure states.
Analog quantum sensing protocols are typically defined with a (generally time-dependent) generator of time evolution known as a \emph{Hamiltonian}.

More generally, the state of an open quantum system, that is, a quantum system that interacts with an environment, can be described by a statistical mixture of pure states, known as a \emph{mixed state} or \emph{density operator}, which is represented by a complex-valued $2^n \times 2^n$ matrix.
Under common assumptions about the environment of an open quantum system, most importantly, Markovianity, the time evolution of a mixed state is governed by a system of ordinary differential equations (ODEs) that involve the Hamiltonian and additional data to describe system-environment interactions.
Crucially, all objects involved in the description of a quantum system and its time evolution are generally complex-valued.

\subsection{Quantum sensing and classical simulation}
\label{sec:quantumsensing}
Pioneering advances in the generation and control of entangled states in noisy intermediate-scale quantum (NISQ) devices are opening opportunities for achieving major quantum enhancements to state-of-the-art tabletop metrological devices. Applications range from time and frequency standards to searches for physics beyond the Standard Model. A key metric for any quantum-enhanced metrological (sensing) device is the quantum Fisher information (QFI), which quantifies the sensitivity of a system to perturbations whose magnitude is characterized by a classical parameter of interest. Classically, estimating a physical parameter with $n$ independent sensors improves the achievable relative precision (equivalently, the inverse of the Fisher information) by $O(n^{-1/2})$. However, exploiting quantum mechanics to entangle the sensors enables this precision to be improved up to the so-called Heisenberg limit of $O(n^{-1})$.

We are developing a toolbox to design variational quantum circuits (VQCs) that can be used to prepare optimal states for quantum metrology in noisy settings, that is, in the context of open quantum systems, which is an open problem in quantum science. The task is to prepare a quantum state that is described by the density operator $\rho(x)$ and obtain the optimal parameters $x$ that determine a sequence of logical operations in the VQC. Preparing a general quantum state $\rho(x)$ with a noisy VQC and computing the associated QFI requires a resource-intensive solution of a complex-valued initial value problem and the computation of the spectral decomposition or diagonalization of the resulting density operator.
It is thus crucial to design efficient optimization methods that minimize the number of calls to the calculation of the QFI.

The quantum state prepared by a noisy VQC can be obtained by solving the Lindblad master equation,
\begin{align}
  \dot{\rho}  = -i [H,\rho]_-
  + \sum_j \gamma_j \left(J_j \rho J_j^\dag
  - \frac12 \left[J_j^\dag J_j,\rho\right]_+ \right).
  \label{eq:master_eq}
\end{align}
Here $\dot{\rho}$ is the time derivative of the state $\rho$, $i$ is the imaginary unit, $[A,B]_\pm = AB \pm BA$ is a commutator ($-$) or anti-commutator ($+$), $j$ indexes particular dissipation (noise) processes, $\gamma_j$ is a positive real dissipation rate, $J_j$ is a \emph{jump operator} represented by a complex-valued matrix describing dissipation, and $J_j^\dag$ is the Hermitian adjoint (complex conjugate and transpose) of $J_j$.

The Hamiltonian $H$ captures dynamics arising from processes internal to the quantum system, as well as (coherent) external controls, for example, for applying quantum logic gates.
The Hamiltonian thereby generally depends both on time $t$ and on a collection of parameters $x$ that encode a VQC, $H = H(t, x)$.
The jump operators $J_j$ describe the loss of information from the quantum system, for example due to undesired couplings to the external environment (such as stray electromagnetic fields or auxiliary quantum systems).
These jump operators, and the corresponding dissipation rates $\gamma_j$, are therefore typically fixed and independent of time.
In the context of NISQ platforms, the master equation  rarely admits an exact or even approximate analytic solution and is therefore solved by numerical integration.
Operationally, this is achieved by writing the master equation as a system of coupled ODEs of the form $\dot{\rho} = \mathcal{L}(\rho)$, where the \emph{Liouvillian} $\mathcal{L}$ is a linear operator on the space of density operators.

Once $\rho(x)$ has been obtained by numerical integration of the master equation, the associated QFI is computed through matrix diagonalization. Specifically, the QFI is given by 
\begin{align} \label{eq:qfi}
  \displaystyle F(x) = \sum_{i=1}^n \sum_{j < i} \frac{(\lambda_i(x) - \lambda_j(x))^2}{\lambda_i(x) + \lambda_j(x)}
  \left| \braket{\psi_i(x) | G | \psi_j(x)} \right|^2,
\end{align}
where $\{\lambda_i, \psi_i\}$ denote the eigenvalue-eigenvector pairs of $\rho(x)$ and the summation over $i$ and $j$ implicitly ignores terms where $\lambda_i(x) + \lambda_j(x) = 0$. The QFI captures the fundamental precision with which a parameter $\theta$ can be estimated when it is imprinted onto $\rho(x)$ via the unitary transformation $\rho(x) \to e^{-i\theta G} \rho(x) e^{i\theta G}$, and serves as our ultimate figure of merit (reward function) in determining the optimal state $\rho(x)$ for quantum metrology. The complex-valued numerical integration procedure described above is hence a subroutine to evaluate $F(x)$. Alternative expressions to \eqref{eq:qfi} can be obtained \cite{Fiderer2021qfi,Rezakhani2019qfi}, although all require operations with overhead similar to what is needed to diagonalize $\rho(x)$. Thus, the high cost of evaluating $F(x)$ motivates the use of automatic differentiation (autodiff) for gradient-based methods to optimize $F$ with respect to $x$.

\subsection{Existing quantum computing software}
Several popular tools for quantum computing exist.
\texttt{PennyLane} is a cross-platform Python library for differentiable programming of quantum computers~\cite{bergholm2022pennylane}.
It allows for the expression of quantum circuits and the optimization of  hybrid classical-quantum models using interfaces to \texttt{autograd}, \texttt{PyTorch}, \texttt{TensorFlow}, and \texttt{JAX}.
\texttt{Cirq} and \texttt{Qiskit} are also open-source software development kits for representing and manipulating quantum programs at the level of pulses, circuits, and application modules~\cite{cirq_developers_2022_7465577, Qiskit}.
However, circuit-level simulators are not suitable for the QFI optimization that we are performing, because of their overly simplified noise models.
Specifically, circuit-level noise models typically treat coherent (Hamiltonian) and incoherent (dissipative) dynamics as distinct processes separated in time.
While this treatment of noise may be appropriate in the low-noise regime, or for understanding certain qualitative features of quantum programs, we wish to consider moderate-to-high levels of dissipation with noise models that more faithfully represent the underlying physics, namely, with quantum dynamics described by Eq.~\eqref{eq:master_eq}.

In principle, circuit-level simulators, including the above, are capable of representing arbitrary quantum processes through the construction of quantum channels.
However, constructing a quantum channel corresponding to the dynamics in Eq.~\eqref{eq:master_eq} requires constructing a $4^n\times 4^n$ matrix representation of the Liouvillian $\mathcal{L}$ for $n$ qubits, and exponentiating this matrix to compute the time evolution operator $e^{t\mathcal{L}}$.
In the case of a time-dependent Liouvillian, exponentiating as $e^{t\mathcal{L}}$ is replaced by a numerical evaluation of the product integral $\prod_t e^{\mathrm{d}t\mathcal{L}}$.
This task is considerably more expensive than numerically integrating a $2^n\times 2^n$ density operator.
In any case, this construction can have the side effect of trivializing the circuit representation of a quantum program by reducing it to a single operation, or a series of operations, that addresses all qubits.

\texttt{QuTiP} is another widely used open-source Python library for simulating the dynamics of open quantum systems~\cite{JOHANSSON20121760, JOHANSSON20131234}.
\texttt{QuTiP} provides numerical simulations of a wide variety of Hamiltonians and Liouvillians, including those with arbitrary time dependence, commonly found in a wide range of physics applications such as quantum optics, trapped ions, superconducting circuits, and quantum nanomechanical resonators.
Although \texttt{QuTiP} has all of the tools we require for numerical integration, it is not suitable for our work because these tools are not written in a manner that is compatible with automatic differentiation.

\subsection{Our approach}

Because we are unable to use existing tools, we explore implementing our code within languages and frameworks that support differentiable programming. The first step in our computation is the numerical integration of a system of ODEs defined by an initial value problem of the form
\begin{align}
  \frac{dy}{dt} = f(t,y),
  &&
  y(t_0) = y_0,
\end{align}
where $f$ is an arbitrary function of real-valued time $t$ and a complex-valued vector $y$ and ($t_0, y_0$) are the initial value data. The second step in our computation is the eigendecomposition of a real-valued matrix. Both these steps are implicit functions that should not be differentiated naively. Their derivatives should instead be formulated on a higher abstraction level~\cite{hueckelheim2023a}. 

While several frameworks provide the capability to solve an ODE with floating-point data, solving an ODE with complex-valued data is typically not supported.
Even when frameworks do support solving complex-valued ODEs, they do not support differentiating the solution to the ODE. While frameworks  support eigendecomposition in general, they differ in whether they can provide derivatives for the eigendecomposition of a nonsymmetric input matrix as well as whether they can provide derivatives with respect to the eigenvectors as well as eigenvalues.

The rest of the paper is organized as follows. \Cref{sec:implementation} describes the different approaches we have examined for solving the complex-valued ODE integration and eigendecomposition. \Cref{sec:otherconsiderations} discusses other considerations, and \Cref{sec:conclusions} presents our conclusions.

\section{Implementation}
\label{sec:implementation}
Our initial implementation was written in Python using \texttt{SciPy}~\cite{2020SciPy-NMeth}. Here we describe that approach and its alternatives.

\subsection{ODE Solver}

An important piece of the computation is a call to \mintinline[breaklines]{python}{scipy.integrate.solve_ivp} to solve an initial value problem for a system of ODEs.
In our computation (see \Cref{lst:scipy}) \mintinline[breaklines]{python}{y0} is
an initial density operator, and so the call results in the time evolution of the density operator for a given time period.

\begin{code}[t]
\begin{tcolorbox}[left=20pt,title=State Evolution in \texttt{SciPy}]
\begin{minted}{python}
def evolve_state(
    density_op: np.ndarray,
    time: float,
    hamiltonian: np.ndarray,
) -> np.ndarray:
    solution = scipy.integrate.solve_ivp(
        time_deriv,
        (0, time),
        density_op.ravel(),
        t_eval=[time],
        args=(hamiltonian,),
        method="DOP853",
    )
    final_vec = solution.y[:, -1]
    return final_vec.reshape(density_op.shape)
\end{minted}
\end{tcolorbox}
\caption{
Python code to time-evolve a density operator using \mintinline[breaklines]{python}{scipy.integrate.solve_ivp}.
Here \texttt{time\_deriv} is a method that accepts a time $t$, density operator $\rho$, and Hamiltonian $H$ as arguments and returns the time derivative $\dot\rho = \texttt{time\_deriv}(t, \rho, H)$.
Note that the Hamiltonian may itself be a function of time, which the \texttt{time\_deriv} method would need to account for.
}
\label{lst:scipy}
\end{code}

The density operator is a complex-valued matrix, and time is a real-valued scalar.
The third argument to the solver, \mintinline[breaklines]{python}{hamiltonian}, is a complex-valued matrix as well.
\mintinline[breaklines]{python}{scipy.integrate.solve_ivp} supports both sparsity and complex-valued inputs.
\texttt{SciPy} does not, however, have an in-built autodiff capability, which we need for gradient-based optimization. If we use the \texttt{autograd}~\cite{maclaurinautograd} package to compute derivatives, we will be limited to \mintinline[breaklines]{python}{scipy.integrate.odeint}, which uses LSODA~\cite{10.1137/0904010} from the FORTRAN library ODEPACK~\cite{osti_145724}, whereas \mintinline[breaklines]{python}{solve_ivp} provides several methods. More important, \mintinline[breaklines]{python}{odeint} fails with the following error:

\mintinline[breaklines]{bash} {Cannot cast array data from dtype('complex128') to dtype('float64') according to the rule 'safe'} ,

\noindent which implies that it does not support complex-valued inputs.

\subsubsection{TensorFlow}
\texttt{TensorFlow Probability} provides the capability to numerically solve ordinary ODEs with floating-point initial values through the \mintinline[breaklines]{python}{tfp.math.ode.BDF} class~\cite{dillon2017tensorflow}. Running the primal code with complex initial values requires one to provide a Jacobian function to the solver because of the following limitation:

\mintinline[breaklines]{bash}{NotImplementedError: The BDF solver does not support automatic Jacobian computations for complex dtypes.}

\begin{code}[h!]
\begin{tcolorbox}[left=20pt,title=State Evolution in \texttt{TensorFlow}]
\begin{minted}{python}
def evolve_state(
    density_op: tf.Tensor,
    time: float,
    hamiltonian: tf.Tensor | tf.sparse.SparseTensor,
    rtol: float = 1e-8,
    atol: float = 1e-8,
) -> tf.Tensor:
    if time == 0:
        return density_op
    if time < 0:
        time, hamiltonian = -time, -hamiltonian
    result = tfp.math.ode.BDF(
        rtol=rtol,
        atol=atol,
    ).solve(
        time_deriv,
        0,
        density_op,
        solution_times=[0, time],
        jacobian_fn=time_deriv_jac,
    )
    return tf.reshape(result.states[-1], d
                      ensity_op.shape)
\end{minted}
\end{tcolorbox}
\caption{Corollary to the code in \Cref{lst:scipy}, for computing a state $\rho(x)$ using the \texttt{TensorFlow} framework.}
\label{lst:tf}
\end{code}

Otherwise, it provides the same output as the \texttt{SciPy} version (see \Cref{lst:tf}).
Differentiating the code, however, results in an error:

\mintinline[breaklines]{bash}{NotImplementedError: The adjoint sensitivity method does not support complex dtypes.}

\subsubsection{torchdiffeq}
\texttt{torchdiffeq} provides ODE solvers implemented in \texttt{PyTorch}~\cite{torchdiffeq}. Backpropagation through ODE solutions is supported using the adjoint method for constant memory cost.
It is possible to remove a check within \texttt{torchdiffeq} and allow complex initial value arguments to \mintinline[breaklines]{python}{torchdiffeq.odeint}. When this is done, the primal output matches the \texttt{SciPy} output (see \cref{lst:pt}).
\texttt{torchdiffeq} does not, however, return derivatives with respect to the \texttt{time} parameter passed to the numerical integrator.

\begin{code}[t]
\begin{tcolorbox}[left=20pt,title=State Evolution in torchdiffeq]
\begin{minted}{python}
import torch
from torchdiffeq import odeint as odeint

def evolve_state(
    density_op: torch.tensor,
    time: float,
    hamiltonian: torch.tensor,
    rtol: float = 1e-8,
    atol: float = 1e-8,
) -> torch.tensor:
    if time == 0:
        return density_op
    if time < 0:
        time, hamiltonian = -time, -hamiltonian

    t = torch.linspace(0.0, time, 100)
    result = odeint(time_deriv,
                    density_op,
                    t,
                    method="dopri5",
                    rtol=rtol,
                    atol=atol)
    return result[-1]
\end{minted}
\end{tcolorbox}
\caption{Corollary to the code in \Cref{lst:scipy}, for computing a state $\rho(x)$ using the \texttt{torchdiffeq} library.}
\label{lst:pt}
\end{code}

\subsubsection{JAX}
\texttt{JAX} provides \mintinline[breaklines]{python}{jax.experimental.ode.odeint} for integrating systems of ODEs~\cite{jax2018github} (see Listing~\ref{lst:jax}). For computing adjoints, \texttt{JAX} provides an implementation of the approach presented in~\cite{chen2019neural}.
This implementation provides derivatives with respect to the initial state, time, and dictionary arguments such as the Hamiltonian. For complex-valued differentiation, the user must assert that the function is holomorphic. We find that for \mintinline[breaklines]{python}{jax.experimental.ode.odeint}, however, that the complex component of the derivatives are incorrect. It is not possible to test forward mode differentiation of  \mintinline[breaklines]{python}{jax.experimental.ode.odeint} because \texttt{JAX} only provides custom reverse-mode derivatives for \mintinline[breaklines]{python}{jax.experimental.ode.odeint}.

\begin{code}
\begin{tcolorbox}[left=20pt,title=State Evolution in \texttt{JAX}]
\begin{minted}{python}
def evolve_state(
    density_op: jnp.ndarray,
    time: float,
    hamiltonian: jnp.ndarray,
):
    times = jnp.linspace(0.0, time, 2)
    solution = odeint(
        time_deriv,
        density_op,
        times,
        (hamiltonian,),
    )
    return solution[-1]
\end{minted}
\end{tcolorbox}
\caption{
Corollary to the code in \Cref{lst:scipy}, for computing a state $\rho(x)$ using the \texttt{JAX} framework.
}
\label{lst:jax}
\end{code}

\subsubsection{tfdiffeq}
\texttt{tfdiffeq} provides ODE solvers implemented in TensorFlow Eager~\cite{tfdiffeq}. It does not currently provide adjoint capability and therefore does not meet our requirements for gradient-based optimization.

\subsubsection{NeuroDiffEq}
\texttt{NeuroDiffEq} is a package built with \texttt{PyTorch} that uses artificial neural networks
to solve ODEs and PDEs~\cite{chen2020neurodiffeq}. The package provides \mintinline[breaklines]{python}{neurodiffeq.conditions.IVP} to specify an initial value problem.
Setting the initial value to be complex-valued, however, results in an error.

\subsubsection{DeepXDE}
\texttt{DeepXDE} is a deep-learning library for solving differential equations~\cite{doi:10.1137/19M1274067}. It can solve forward problems given initial and boundary conditions, as well as inverse problems given some extra measurements.
Setting the initial value to be complex-valued results in an error.

\subsubsection{PyDEns}
\texttt{PyDEns} is a framework for solving ODEs and PDEs using neural networks~\cite{alex2019pydens}. Setting the initial value to be complex-valued results in an error.

\subsubsection{DifferentialEquations.jl}
\texttt{DifferentialEquations.jl} is a suite for numerically solving differential equations written in Julia and available for use in Julia, Python, and R~\cite{rackauckas2017differentialequations}. It provides Julia implementations of solvers for various differential equations including ODEs. This suite can solve ODEs with complex-valued initial values. \texttt{SciMLSensitivity.jl} is the automatic differentiation (AD) and adjoints system for \texttt{DifferentialEquations.jl}~\cite{rackauckas2020universal}. It provides automatic differentiation overloads for improving the performance and flexibility of AD calls over the ODE solver. It is compatible with several AD libraries including \texttt{Zygote.jl}~\cite{innes2019dont}. Differentiating the solver with a complex-valued initial value problem, expressed using \texttt{SciMLSensitivity.jl} and \texttt{Zygote.jl}, however, leads to an error:

\mintinline[breaklines]{bash}{ERROR: LoadError: Output is complex, so the gradient is not defined.}

\subsubsection{Diffrax}
\texttt{Diffrax} is a \texttt{JAX}-based library providing numerical differential equation solvers~\cite{kidger2021on}. It provides ordinary, stochastic, and controlled differential equation solvers and several methods including \texttt{Tsit5}, \texttt{Dopri8}, symplectic solvers, and implicit solvers. It provides both forward-mode and backward-mode differentiation. 

\begin{code}[t]
\begin{tcolorbox}[left=20pt,title=State Evolution in \texttt{Diffrax}]
\begin{minted}{python}
def evolve_state(
    density_op: np.ndarray,
    time: float,
    hamiltonian: np.ndarray,
    solver: diffrax.AbstractSolver,
    **diffrax_kwargs: object,
) -> np.ndarray:
    diffrax_kwargs["dt0"] = 0.002
    def _time_deriv(time: float, 
                    density_op: np.ndarray) -> np.ndarray:
        return time_deriv(time, density_op)

    term = diffrax.ODETerm(_time_deriv)
    solver = diffrax.Tsit5()
    solver_args = dict(t0=0.0, 
                       t1=time.real, 
                       y0=density_op, 
                       args=(hamiltonian,))
    diffrax_kwargs["max_steps"] = diffrax_kwargs.get(
        "max_steps", None)
    solver_args |= dict(adjoint=diffrax.DirectAdjoint())
    solution = diffrax.diffeqsolve(term,
                                    solver, 
                                    **solver_args, 
                                    **diffrax_kwargs)
    return solution.ys[-1]
\end{minted}
\end{tcolorbox}
\caption{
Corollary to the code in \Cref{lst:scipy}, for computing a state $\rho(x)$ using the \texttt{Diffrax} library.
This code is compatible with both complex values and automatic differentiation. 
}
\label{lst:diffrax}
\end{code}

Diffrax experimentally supports the complex-valued IVP. Because Diffrax is built using JAX it is differentiable. We find in practice though that the results obtained from calling \texttt{jax.jacfwd} for our complex-valued IVP are correct and the results obtained from calling \texttt{jax.jacrev} are incorrect.

\subsection{Eigenvalue Decomposition}
The expression for QFI in \eqref{eq:qfi} is a function of eigenvalues and eigenvectors of a complex-valued density matrix. Thus, differentiation of QFI with respect to parameters $x$ requires differentiation of the eigenvalue decomposition operation. 

Computing the derivatives of eigenvalues and eigenvectors has a long history~\cite{Durbha1620260202}.
The rules for differentiating an eigenvalue decomposition of a real-valued matrix have been documented in ~\cite{Giles2008AnEC}.
These rules have been extended via the complex-valued chain rule for computing the derivative of the operation with complex-valued eigenvectors~\cite{Boeddeker2017}. We have examined this approaches for our complex-valued density matrix.

Consider the eigenvalue problem $\rho(x) \psi = \lambda \psi.$
We will introduce an additional constraint to the eigenvalue problem, $\braket{\psi_i | \psi_i} = 1$, to fill up a remaining degree of freedom. 
Naively, one can compute the partial derivatives $\partial_k \psi$ and $\partial_k \lambda$ through a simple application of the chain rule, yielding
\begin{equation}
\label{eq:matrix_ders}
\left[
\begin{array}{cc}
\rho(x) - \lambda I & -\psi\\
\bar{\psi}^\top & 0\\
\end{array}
\right]
\left[
\begin{array}{c}
\partial_k \psi\\
\partial_k \lambda
\end{array}
\right]
=
\left[
\begin{array}{c}
-\partial_k \rho(x) \psi\\
0
\end{array}
\right].
\end{equation}
The solution of \eqref{eq:matrix_ders} is evidently only unique (well-defined) provided the Hermitian matrix on the left-hand side is nonsingular. 
It is thus also evident from \eqref{eq:matrix_ders} that well-definedness of derivatives fails whenever $\rho(x)$ has an eigenvalue of multiplicity greater than one. 
This observation also has implications for the stability of derivative computations via this naive approach whenever two eigenvalues are nearly equal. 

After some experimentation, we determined that the eigenvalues and eigenvectors of density matrices $\rho(x)$ exhibit interesting properties as a function of $x$. 
\matt{Need Michael or a similar expert to chime in here about why any of this paragraph might be true.}
In particular, these density matrices always have multiplicity.
Moreover, we observe that if one samples a random parameterization $x$, then with high probability, the eigenvalue functions corresponding to an eigenvalue of $\rho(x)$ with multiplicity appear identical on neighborhoods of $x$. 

We explored the following frameworks for computing the derivatives we required and settled upon using hand-coded derivatives.

\subsubsection{JAX}
\texttt{JAX} provides \mintinline[breaklines]{python}{jax.numpy.linalg.eig} and \mintinline[breaklines]{python}{jax.numpy.linalg.eigh} for computing the eigenvalues and eigenvectors of a square array and a Hermitian matrix respectively. Derivatives for \mintinline[breaklines]{python}{jax.numpy.linalg.eigh} with respect to both the eigenvalues and eigenvetors are computed using the approach from~\cite{Boeddeker2017} assuming distinct eigenvalues.

\subsubsection{Pytorch}

\texttt{Pytorch}'s
\mintinline[breaklines]{python}{torch.linalg.eigh} computes the eigenvalue decomposition of a complex Hermitian or real symmetric matrix. \mintinline[breaklines]{python}{torch.linalg.eig} computes the eigenvalue decomposition of a square matrix when all its eigenvalues are different. Derivatives computed with respect to the eigenvectors are finite when the input matrix has distinct eigenvalues. Furthermore, if the distance between any two eigenvalues is close to zero, the derivatives are numerically unstable. For the input matrix in our problem, consequently, we observe the following error:

\mintinline[breaklines]{text}{RuntimeError: linalg_eigh_backward: The eigenvectors in the complex case are specified up to multiplication by e^{i phi}. The specified loss function depends on this quantity, so it is ill-defined.}

\subsubsection{TensorFlow}
\mintinline[breaklines]{python}{tf.linalg.eig} computes the eigendecomposition of a batch of matrices and \mintinline[breaklines]{python}{tf.linalg.eigh} computes the computes the eigendecomposition of a batch of self-adjoint matrices. Reverse-mode derivatives with respect to the eigenvalues and eigenvectors are available using the approach from~\cite{Boeddeker2017}. Here too the gradient becomes
infinite when eigenvalues are not unique.

\subsubsection{Handcoded Derivatives}
We employed a technique for approximate computation of derivatives of eigenvalues and eigenvectors outlined in \cite[Section 4.2]{chu1990multiple}. 
This particular technique acknowledges that general eigenvalue problems are nondifferentiable when an eigenvalue of $\rho(x)$ has multiplicity and elects to compute the derivatives of each invariant subspace associated with eigenvectors of multiplicity greater than one, and computes the derivative of the associated \emph{average} of the eigenvalue with multiplicity. 
This averaging of eigenvalues in a neighborhood of $\rho(x)$ admits a solution (approximate derivative) that is always well-defined. 
By our empirical observations, this average derivative of eigenvalues is, with very high probability, the \emph{exact} derivative of the common eigenvalues.

\section{Other considerations}
\label{sec:otherconsiderations}
While we have focused on support for the automatic differentiation of complex-valued initial value problems, classical simulations of quantum dynamics pose additional challenges that will need to be addressed going forward.
One such challenge is the support of sparse matrix types.
Physically relevant Hamiltonians and Liouvillians are generally sparse, which significantly reduces both the memory and time complexity of evaluating the time derivative $\dot\rho = \mathcal{L}(\rho)$.
While several frameworks support sparse matrix types, these types are not compatible with the corresponding implementations of autodiff-compatible ODE solvers.
For example, \texttt{PyTorch} allows conversion to and from sparse representations, but these are not compatible, at the moment, with \texttt{torchdiffeq}'s implementation of \texttt{odeint}.

A second challenge comes from the fact that quantum systems have an exponentially large state space.
Implementations of the derivative of an ODE solver, such as those in \texttt{JAX}, store the state of the solution at every time step to compute the adjoint at that time step.
Because density operators have an exponentially large  number of qubits in a quantum system, storage requirements can quickly overcome available memory even for small (e.g., $n=6$) qubit numbers.
Therefore, techniques such as checkpointing, which are already supported by many frameworks, should be combined with implementations of the ODE solver. We found that only \texttt{Diffrax} supports this approach.

\section{Conclusions}
\label{sec:conclusions}
We have examined different implementations of a differentiable ODE solver with complex-valued initial conditions for gradient-based optimization. We have also examined the availability of a suitable differentiable eigendecomposition operation.
With increasing timeliness and a growing interest in quantum technologies, popular frameworks, and packages should expect that they may be used for developing quantum applications and should be enhanced to support complex-valued data types.
Wherever support is lacking, these frameworks should also be enhanced to support sparse matrices, which are ubiquitous in quantum applications.

\ifdefined\isaccepted
\input{ack.tex}
\fi

\bibliography{refs}
\bibliographystyle{siam}

\onecolumn
\framebox{\parbox{\columnwidth}{The submitted manuscript has been created by UChicago Argonne, LLC, Operator of Argonne National Laboratory (`Argonne'). Argonne, a U.S. Department of Energy Office of Science laboratory, is operated under Contract No. DE-AC02-06CH11357. The U.S. Government retains for itself, and others acting on its behalf, a paid-up nonexclusive, irrevocable worldwide license in said article to reproduce, prepare derivative works, distribute copies to the public, and perform publicly and display publicly, by or on behalf of the Government.  The Department of Energy will provide public access to these results of federally sponsored research in accordance with the DOE Public Access Plan. \url{http://energy.gov/downloads/doe-public-access-plan}.}}
\end{document}

%% file: ack.tex
\section*{Acknowledgements}
This work was supported by the U.S. Department of Energy, Office of Science, Advanced Scientific Computing Research, Exploratory Research for Extreme-Scale Science program.
This material is based upon work supported by the Applied Mathematics activity within the U.S. Department of Energy, Office of Science, Office of Advanced Scientific Computing Research, under contract DE-AC02-06CH11357.
This material is based upon work supported by the U.S. Department of Energy, Office of Science, Office of Advanced Scientific Computing Research, under the Accelerated Research in Quantum Computing and Applied Mathematics programs, under contract DE-AC02-06CH11357.

We thank Paul Hovland,  Michel Schanen, and Daniel Adrian Maldonado for insightful comments. We thank Ricky Tian Qi Chen for modifying \texttt{torchdiffeq} to handle complex initial values. We thank Chris Rackaukas for suggesting \texttt{DifferentialEquations.jl}.